\documentclass[9pt,aps,prl,reprint,floatfix,superscriptaddress]{revtex4-1}

\usepackage{cancel}
\usepackage[english]{babel}
\usepackage{multirow}
\usepackage{graphicx}
\usepackage{dcolumn}
\usepackage{bm}
\usepackage{chemfig}
\usepackage{amsmath}
\usepackage{amsfonts,amssymb}
\usepackage[colorlinks=true,linkcolor=blue]{hyperref}
\usepackage[shortlabels]{enumitem}
\usepackage{braket}
\usepackage{adjustbox}
\usepackage{xcolor}
\usepackage{gensymb}

\begin{document}

\title{Molecules in Environments: Towards Systematic Quantum Embedding of Electrons and Drude Oscillators}

\author{Matej Ditte}
\email[]{matej.ditte@uni.lu}
\affiliation{Department of Physics and Materials Science, University of Luxembourg, L-1511 Luxembourg City, Luxembourg}

\author{Matteo Barborini}
\affiliation{Department of Physics and Materials Science, University of Luxembourg, L-1511 Luxembourg City, Luxembourg}

\author{Leonardo Medrano Sandonas}
\affiliation{Department of Physics and Materials Science, University of Luxembourg, L-1511 Luxembourg City, Luxembourg}

\author{Alexandre Tkatchenko}
\email[Corresponding author:]{alexandre.tkatchenko@uni.lu}
\affiliation{Department of Physics and Materials Science, University of Luxembourg, L-1511 Luxembourg City, Luxembourg}

\date{March 28, 2023}

\begin{abstract}
We develop a quantum embedding method that enables accurate and efficient treatment of interactions between molecules and an environment, while explicitly including many-body correlations. 
The molecule is composed of classical nuclei and quantum electrons, whereas the environment is modeled \textit{via} charged quantum harmonic oscillators. 
We construct a general Hamiltonian and introduce a variational ansatz for the correlated ground state of the fully interacting molecule/environment system. 
This wavefunction is optimized via variational Monte Carlo and the ground state energy is subsequently estimated through diffusion Monte Carlo. The proposed scheme allows an explicit many-body treatment of electrostatic, polarization, and dispersion interactions between the molecule and the environment. We study solvation energies and excitation energies of benzene derivatives, obtaining excellent agreement with explicit \textit{ab initio} calculations and experiment.
\end{abstract}

\maketitle

The challenges of computational electronic-structure methods have evolved from the precise description of short-range interactions involving few electrons to the accurate modeling of longer-range many-electron effects for molecules and materials~\cite{Sun2016,Jones2020,Tkatchenko2017rev}.
This shift has been enabled by the enormous progress in hardware capabilities combined with the development of new effective physical models and numerical techniques. 

In particular, embedding approaches, such as the hybrid QM/MM~\cite{Fersht2013,Khare2007,Torras2009,Zimmerman2011,Nasluzov2001,Nasluzov2003,Brandle1998,Guareschi2016,Senn2009,Acevedo2010,Huang2017,Alves2015,Silva2015,Bao2019,Marin2019,Robertazzi2006,Gkionis2009,Kupfer2014,Osoegawa2019,loc+23pho,Cho2010} or QM/QM strategies~\cite{Kotliar2006,Lan2015,Knizia2012,Wesolowski2015,Libisch2014,Manby2012,Katin2017,Friedrich2007,Seth2002,Vogiatzis2015,Huo2016,Lee2019,Coughtrie2018,Neugebauer2005,Jacob2006,Exner2012,Exner2004,Bockstedte2018,Lau2021,Schafer2021,Ma2021,Petocchi2020,Yeh2021,Nowadnick2015,Chen2022,Nusspickel2022}, based on the partitioning of the system of interest in different subspaces, each treated with a different level of theory and thus of accuracy~\cite{Sun2016,Jones2020}, have been devised to study excitation energies, solvation energies or charge transfer between fragments in a complex environment~\cite{bov+13acie,Torras2009,Guareschi2016,Marin2019,Wesolowski2015,Neugebauer2005,Schafer2021,Cho2010}. 

Even though embedding approaches can significantly improve the computational efficiency, this comes at the cost of approximating collective interactions within and between the subsystems, thus excluding important many-body effects in the overall description~\cite{Sun2016,Jones2020}.
Furthermore, systems dominated by non-covalent interactions~\cite{Benali2014,AlHamdani2021} or by static correlations~\cite{ret+14jctc,boy+21jcp,bao+16jctc,Varsano2017,bar+22prb,poz+13jacs,bar+15jctc,Abe2013} remain challenging, since they require high-level \emph{ab initio} methods, such as coupled cluster (CC)~\cite{Bartlett2007} or quantum Monte Carlo (QMC)~\cite{fou+01rmp,kal+08ch8,bec+17}, for which systematic convergence can be guaranteed only for relatively small systems~\cite{Benali2014,AlHamdani2021}.

\begin{figure}[t]
\centering
\includegraphics[width=0.6\columnwidth]{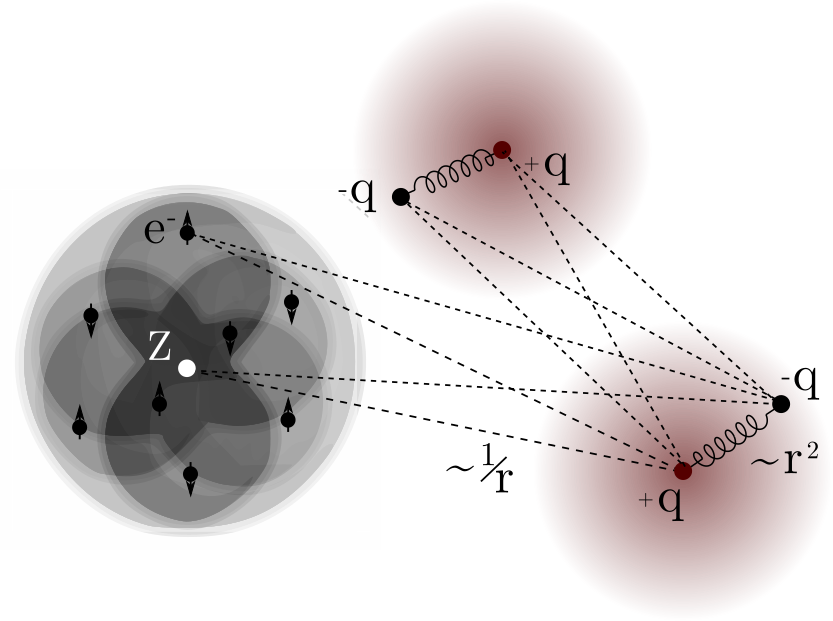}
\caption{{\bf Schematic representation of the embedding model for a single atom and two Drude oscillators.} Strings indicate the harmonic interactions between the drudon and their corresponding centers, the dashed lines represent the Coulomb interactions between all other pairs of charged particles (for simplicity, only selected interactions are explicitly shown).}
\label{Fig:interactions}
\end{figure}

In this Letter, we propose a novel approach, based on the embedding of electronic systems (atoms/molecules) within an environment of charged quantum harmonic (Drude) oscillators, which can reproduce the quantum fluctuations responsible for the non-covalent interactions and for the long-range response of realistic molecules and materials (see Fig.~\ref{Fig:interactions})~\cite{Martyna2009,Tkatchenko2012,MartynaPRL2013,MartynaMolPhys2013,Martyna2013,odb+15cpl,odb+16jcp,Martyna2019,Tkatchenko2009,Vaccarelli2021,poi+22jctc,Karimpour2022prr,Karimpour2022jpcl}. Hence, our approach explicitly describes many-body correlations between the subsystem and the environment. 

Although harmonic oscillators have been used in classical polarizable force fields~\cite{lem+16cr,bui+23nl,som+05jpca}, the novelty of our method lies in the quantum description of the full system of atoms and quantum Drude oscillators (QDOs) through one unified Hamiltonian for which we construct a ground state depending on all degrees of freedom. 
Furthermore, the proposed approach can be used with arbitrary \textit{ab initio} methods, including CC~\cite{Bartlett2007} and QMC~\cite{fou+01rmp,kal+08ch8,bec+17}.

Within the Born-Oppenheimer approximation, each QDO consists of a pair of charged particles corresponding respectively to the \text{center of the oscillator}, with charge $q$ and position $\textbf{R}^{O}$ and to a distinguishable quantum particle called drudon of charge -\textit{q}, mass $\mu$ and position $\textbf{r}^d$. 
The drudon interacts with its center via a quadratic potential $V_{\text{QDO}}\left(r\right)=\frac{1}{2}\mu\omega^2 r^2$ that depends on the mass $\mu$ and on the frequency $\omega$ that determine the slope of the quadratic well, and on their relative Euclidean distance $r=|\textbf{r}^d-\textbf{R}^{O}|$. 
For systems containing more QDOs, all pairs of charged particles (except the drudon and its own center) interact via the Coulomb potential.
In such a model the non-covalent interactions, such as dispersion, arise from the quantum fluctuations of the charge densities that are obtained through the coupling of the quantum states of each QDO.
Naturally, each QDO depends on a set of well defined parameters $\{\textbf{R}_{i}^{O},q_{i},\mu_{i},\omega_{i} \}_{i=1}^{N_{d}}$ that are obtained by reproducing the leading order polarizabilities and dispersion coefficients of real matter~\cite{Martyna2013} from accurate \textit{ab initio} or experimental data~\cite{Margoliash1978,Tkatchenko2009,Tkatchenko2012,Vaccarelli2021,Szabolcs2022}.

Once the parameters have been chosen, the analytical expression of the Hamiltonian of a system of $N_d$ interacting QDOs, the \textit{drudonic Hamiltonian}, is written in atomic units as
\begin{multline}
\hat{H}^{d}=
\sum_{i=1}^{N_{d}} \hat{h}_{i}^{d}\left(\textbf{r}_{i}^{d}\right) 
+ 
\sum_{i=1}^{N_{d}} \sum_{j > i}^{N_{d}}
\frac{q_{i}q_{j}}{\left|\textbf{R}_{i}^{O}-\textbf{R}_{j}^{O} \right|}
+\\+ \sum_{i=1}^{N_{d}} \sum_{j > i}^{N_{d}}
\frac{q_{i}q_{j}}{\left|\textbf{r}_{i}^{d}-\textbf{r}_{j}^{d} \right|},
\label{eq:H^O}
\end{multline} 
where $\bar{\textbf{R}}^O=\left ( \textbf{R}_1^O \ldots \textbf{R}_{N_d}^O \right )$ and $\bar{\textbf{r}}^d=\left ( \textbf{r}_1^d \ldots \textbf{r}_{N_d}^d \right )$ are the vectors of the $3N_d$ coordinates respectively of the oscillators' centers and of the drudons, and $\hat{h}_{i}^{d}\left(\textbf{r}_{i}^{d}\right)$ are the single-body operators
\begin{multline}
\hat{h}_{i}^{d}\left(\textbf{r}_{i}^{d}\right) = -\frac{1}{2\mu_{i}}\nabla^{2}_{\textbf{r}_{i}^{d}} + \frac{1}{2}\mu_{i}\omega_{i}^2\left|\textbf{r}_{i}^{d}-\textbf{R}_{i}^{O} \right|^2 - \\ - \sum_{j \neq i}^{N_d}\frac{q_{i}q_{j}}{\left|\textbf{r}_{i}^{d}-\textbf{R}_{j}^{O} \right|},  
\end{multline}
that include the kinetic energy of the $i$th drudon and its interaction energies with the oscillators' centers.

This drudonic Hamiltonian can be used to model the embedding potential that acts on a molecular system containing $N_{n}$ atomic nuclei and $N_{e}$ electrons, described respectively by the coordinate vectors $\bar{\textbf{R}}^{n}$ and $\bar{\textbf{r}}^{e}$. 
The total Hamiltonian of electrons and QDOs (El-QDO) takes the form
\begin{equation}
\hat{H}^{tot}=\hat{H}^{e}+\hat{H}^{d}+\hat{V}_{int}^{d-e}, 
\label{eq:H^tot}
\end{equation}
where $\hat{H}^e$ is the standard electronic Hamiltonian describing the interaction between the electrons and the atomic nuclei (that are parameters of the system) and $\hat{V}_{int}^{d-e}$ is the interaction between the QDOs and the atoms
\begin{multline}
\hat{V}_{int}^{d-e}=
\sum_{i=1}^{N_{e}}\sum_{j=1}^{N_{d}}\left(\frac{q_{j}}{\left|\textbf{r}_{i}^{e}-\textbf{r}_{j}^{d}\right|} - \frac{q_{j}}{\left|\textbf{r}_{i}^{e}-\textbf{R}_{j}^{O}\right|} \right) + \\ 
+ \sum_{i=1}^{N_{n}}\sum_{j=1}^{N_{d}}\left(\frac{Z_{i}q_{j}}{\left|\textbf{R}_{i}^{n}-\textbf{R}_{j}^{O}\right|}- \frac{Z_{i}q_{j}}{\left|\textbf{R}_{i}^{n}-\textbf{r}_{j}^{d}\right|} \right).
\end{multline}
In order to describe molecular systems with static multipole moments of the charge distribution (such as the water molecule) it is necessary to introduce additional point charges in the drudonic Hamiltonian that modify the external potential felt by the drudons and by the electrons. For example in the QDO model of water introduced by Jones \textit{et al.}~\cite{Martyna2019} and used in this work, the authors include in the model three additional point charges (centered on the two hydrogen atoms and near the oxygen atom), see Supplemental Material (SM) for details~\cite{note_SI}. 
These point charges do not interact with the QDOs of the corresponding molecule but only with the QDOs of the other water molecules, and in our case also with the electronic subsystem.
A schematic representation of the interactions that are present in the total Hamiltonian is shown in Fig.~\ref{Fig:interactions}.
As explained below, the QDO model will be additionally corrected by an exchange-repulsion energy term that dominates the short-range interaction region~\cite{Manby2016,Martyna2019,Vaccarelli2021}. 

The ground state of the many-body Hamiltonian defined in Eq.~\eqref{eq:H^tot} is constructed through an appropriate variational ansatz including the explicit many-body correlations between all quantum particles.
In order to integrate and optimize such an ansatz over the set of electronic and drudonic coordinates we choose to work in the framework of QMC methods~\cite{fou+01rmp,Austin2012rev,bec+17}, as implemented in the QMeCha package~\cite{QMeCha}, although the approach can be generalized to other quantum-chemistry methods. 
QMC methods are stochastic integration techniques that are able to compute and minimize the energy functional over a chosen trial wave function (see SM~\cite{note_SI}).
In particular we apply the variational Monte Carlo (VMC) method~\cite{fou+01rmp,Austin2012rev,bec+17}, optimizing the wave function with the stochastic reconfiguration algorithm~\cite{sor+01prb,sor+05prb}. Subsequently, to improve our energy estimations we use the diffusion Monte Carlo (DMC) projection method~\cite{fou+01rmp,Austin2012rev,bec+17} that, within the fixed-node approximation used to solve the electronic sign problem, selects the ground state component of the approximate nodal-surface given by the optimized wave function (see SM~\cite{note_SI}). 
Clearly, in the case of pure QDO systems in which the wave function is always positive and there is no nodal surface, the DMC algorithm converges to the exact solution.

We proceed to build the trial wave function for the system's ground state by expressing it as the product of three contributions
\begin{equation}
\Psi_{tot}=\Psi_{e}(\bar{\textbf{r}}^{e})\Psi_{d}(\bar{\textbf{r}}^{d}) \mathcal{J}_{e-d}(\bar{\textbf{r}}^{e}, \bar{\textbf{r}}^{d}),
\label{eq:psi^tot}
\end{equation}
which are respectively the electronic ansatz $\Psi_{e}(\bar{\textbf{r}}^{e})$, the drudonic one $\Psi_d(\bar{\textbf{r}}_{d})$ and a positive-definite function $\mathcal{J}_{d-e}(\bar{\textbf{r}}^{e}, \bar{\textbf{r}}^{d})$, which contains the correlation effects between the electronic and drudonic systems.. 
In this work, the pure electronic part $\Psi_{e}(\bar{\textbf{r}}^{e})$ is constructed through a Slater determinant of molecular orbitals formed as linear combinations of Gaussian basis sets, multiplied by an electronic Jastrow factor used to explicitly describe the many-body correlation effects between electrons and nuclei (see SM~\cite{note_SI}).

The approximate form of the drudonic wave function $\Psi_d(\bar{\textbf{r}}_{d})$  is chosen to match the exact solution of a system of QDOs interacting through the dipole potential, which can be used to approximate the Coulomb potential at large distances as in the many-body dispersion MBD method~\cite{Tkatchenko2012, Ambrosetti2014}. 
For this reason, it is written as the exponential function of a vector-matrix-vector product
\begin{equation}
\Psi_d(\bar{\textbf{r}}_{d})=\exp\left[-\bar{\textbf{r}}_{dO}^{\top} \textbf{A} \bar{\textbf{r}}_{dO} \right],
\label{eq:psi_d}
\end{equation}
where $\bar{\textbf{r}}_{dO}=\bar{\textbf{r}}^{d} - \bar{\textbf{R}}^{O}$ is the $3N_d$ dimensional vector containing the distances between each drudon and its corresponding center, and $\textbf{A}$ is a (symmetric) coupling matrix containing $3N_{d}\left(3N_{d} +1\right)/2$ independent parameters. 
The coupling matrix explicitly correlates the drudons subject to the field of their charged centers.

Finally, the electron-drudon factor $\mathcal{J}_{e-d}(\bar{\textbf{r}}^{e}, \bar{\textbf{r}}^{d})$ represents the coupling between the two types of quantum systems. 
Again we hypothesize that the interactions between the QDO environment and molecular one are dominated by dipole interactions, thus the ansatz will resemble the form in Eq.~\eqref{eq:psi_d}
\begin{equation}
    \mathcal{J}_{e-d}(\bar{\textbf{r}}^{e}, \bar{\textbf{r}}^{d}) = \exp \left [ \textbf{d}^{\top} \textbf{B} \bar{\textbf{r}}_{dO} \right ], 
\end{equation}
coupling the distance vector $\bar{\textbf{r}}_{dO}$ of the QDO's with the dipole moment of the molecular part
\begin{equation}
    \textbf{d}= \sum_{i=1}^{N_{n}} Z_i\textbf{R}_i^n-\sum_{i=1}^{N_{e}}\textbf{r}_i^e,
\end{equation}
through the rectangular matrix $\textbf{B}$ containing a set of $3 \times 3N_d$ free parameters. We remark that the restriction of the QDO wave function to the dipolar approximation is not a significant limitation, since the variational optimization of the wave function parameters is followed by a stochastic DMC calculation of the many-body energy for both electrons and drudons.

\begin{figure}[!t]
\centering
\includegraphics[width=0.95\columnwidth]{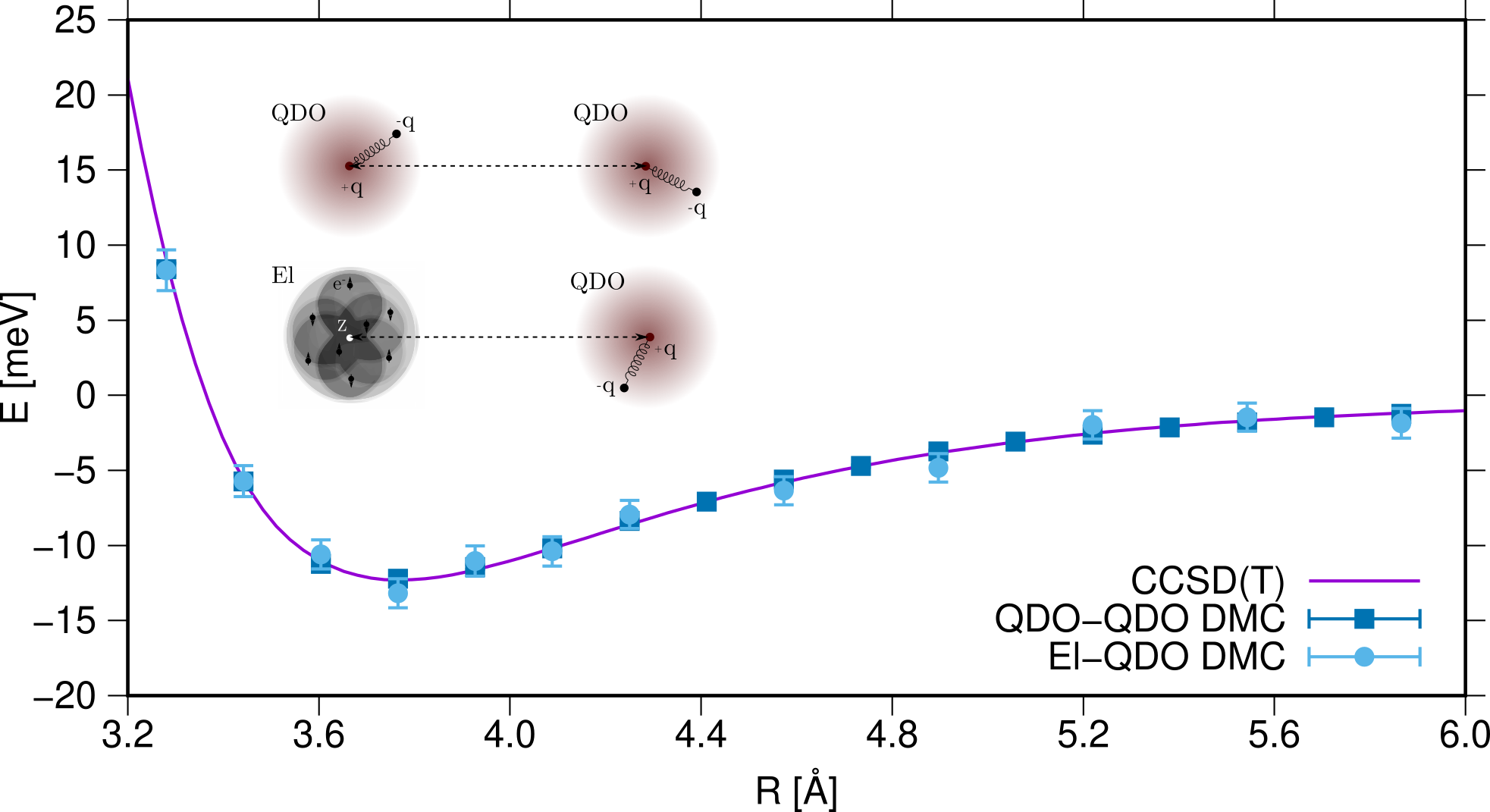}
\caption{Binding energy curve for the Ar dimer obtained using the QDO-QDO model and electrons-QDO (El-QDO) embedding approach (schematically shown in the figure) at the DMC level of theory, with the exponential fit of the short-range repulsion~\cite{note_SI}. The results are compared to the CCSD(T)~\cite{Patkowski2005} reference curve.}\label{Fig:Ar2_fit}
~\\
\includegraphics[width=0.95\columnwidth]{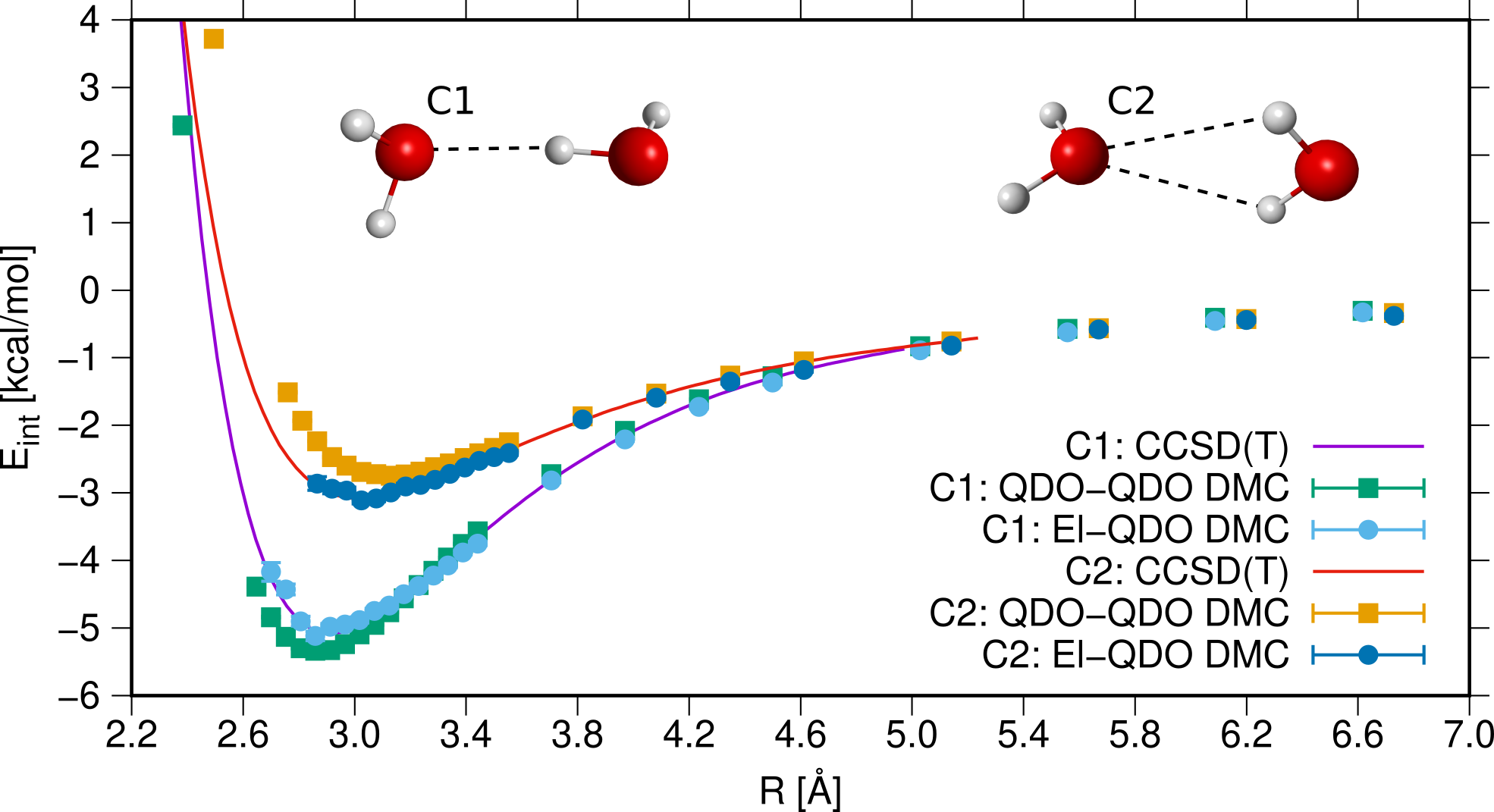}
\caption{Binding energy curve for the water dimer as a function of the oxygen-oxygen distance in the two energetically lowest geometric configurations (C1 and C2)~\cite{Metz2019} obtained using the QDO-QDO and El-QDO (with QDO on the right, approximating the acceptor) models at the DMC level of theory~\cite{note_SI}. The results are compared to the CCSD(T)~\cite{Metz2019} reference curve.}\label{Fig:WA2_fit}
\end{figure}

First, we apply our approach to the Ar~\cite{Patkowski2005} and the water dimers~\cite{Metz2019}. While these systems are small, when taken together they are challenging benchmark examples, requiring an accurate description of exchange, dispersion, polarization, and hydrogen bonding. 
For both Ar and water, a single QDO represents the response of all 8 valence electrons. Accordingly, the QDO model allows a significant reduction in the electronic degrees of freedom. We remark that a bosonic QDO model can be extended to describe exchange~\cite{Fedorov2018,Vaccarelli2021}, but this would require a wavefunction with mixed quantum statistics for electrons and QDOs. Instead, in our current embedding scheme we describe exchange via parametrized short-range exponentials (see SM~\cite{note_SI}). 

In Ar$_2$~\cite{Patkowski2005}, the interaction energy comes from an interplay between correlation in the long range and exchange in the short range.
The attractive dispersion energy is correctly described by the QDO-QDO model and essentially coincides with its counterpart computed from electronic symmetry-adapted perturbation theory (SAPT)~\cite{Sladek2017} (see SM~\cite{note_SI}). Noticeably, the electrons-QDO (El-QDO) embedding model reproduces the full interaction curve upon subtracting the exchange contributions from SAPT, highlighting the ability of our method to capture electrostatic and polarization terms. Together with the corresponding repulsive part, both models match the reference CCSD(T) binding curve~\cite{Patkowski2005}, as shown in Fig.~\ref{Fig:Ar2_fit}.

For the water dimer, where intermolecular interactions are strongly directional, we apply the El-QDO embedding model to study the dissociation energies of its two lowest-energy geometric configurations, C1 and C2, along the O-O axis (Fig.~\ref{Fig:WA2_fit}). 
Here we use the extended QDO model of Martyna \textit{et al.}~\cite{Martyna2019} that contains three point charges and a QDO located at the center-of-mass, substituting the acceptor water molecule (always on the right in Fig.~\ref{Fig:WA2_fit}).
From Fig.~\ref{Fig:WA2_fit} one observes that both C1 and C2 binding curves obtained with the QDO-QDO and El-QDO models are in excellent agreement with the CCSD(T) references~\cite{Metz2019}.
The models are thus able to describe the different interactions when varying the molecular orientation.
Moreover, the comparison of the bare El-QDO interaction curve of the C1 conformer with the SAPT decomposition from Ref.~\citenum{Altun2018} (see SM~\cite{note_SI}), shows that our model is again close to the full binding energy curve, upon subtracting the repulsive exchange contributions, thus correctly capturing both dispersion and polarization effects.
From the atomic and molecular examples of Ar and water dimers, we conclude that our embedding approach is able to accurately describe the manifold of interactions between an electronic subsystem and an embedding environment, in which the latter is described through a combination of QDOs and point charges. 

\begin{figure}[!t]
\centering
\includegraphics[angle=0,width=1.0\columnwidth]{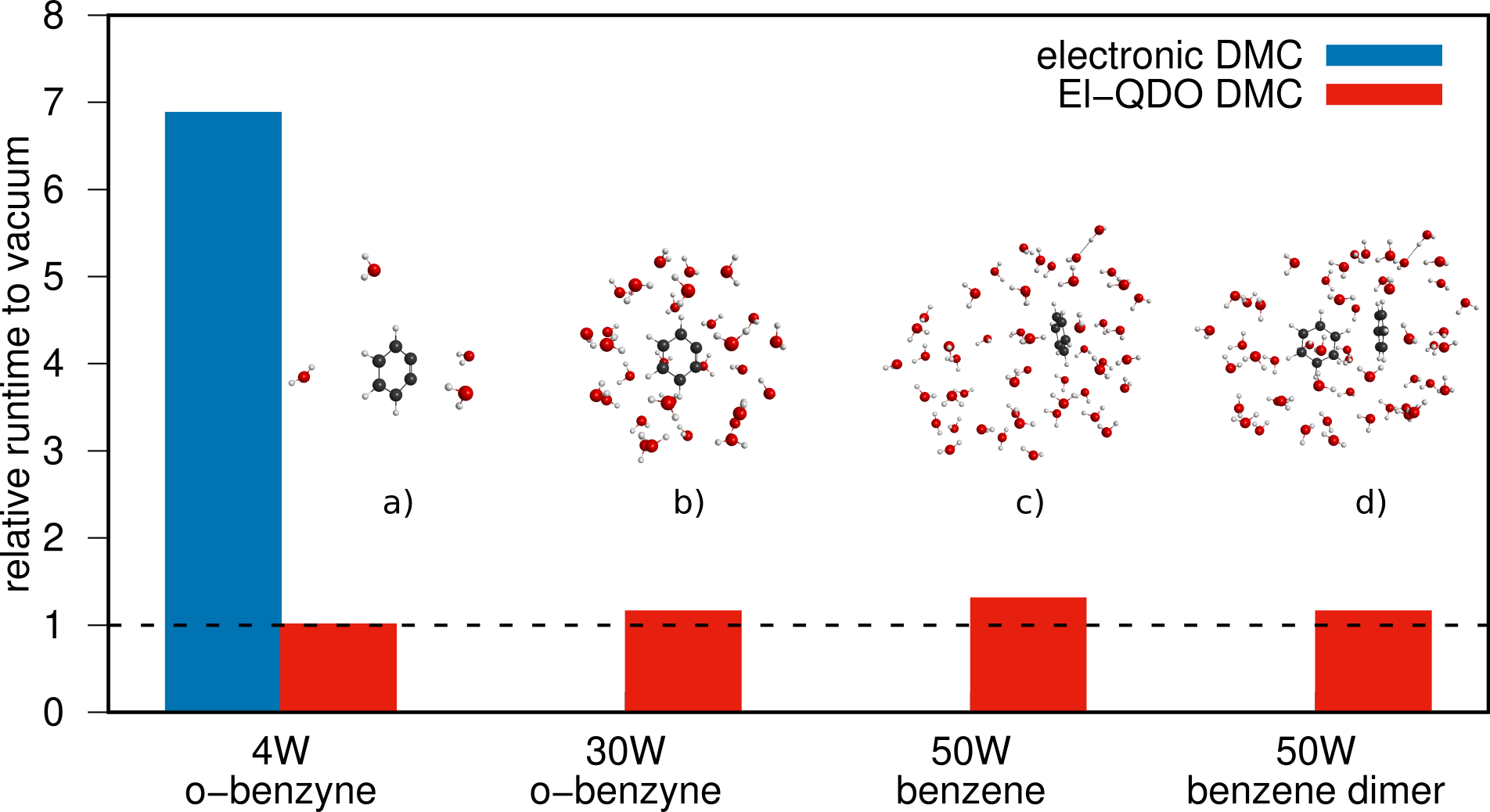}
\caption{Relative runtimes (see SM~\cite{note_SI}) of water-embedded El-QDO DMC calculations with respect to the molecular energies computed with DMC in vacuum. Runtime for electronic DMC is also shown for ortho-benzyne in 4W cage. Inset figures a) to d) correspond to the molecular structures~\cite{Sirianni2022,Shee2019}.}
\label{Fig:runtimes}
\end{figure}

Among the most intriguing applications of embedding approaches are the effects of solvent on the binding energies between molecules and on their electronic excitations~\cite{DiStasio2007,Carter2020}.
Here, we study the binding in T-shaped benzene dimer, prototype for C-H$-\pi$ interactions~\cite{DiStasio2007,Carter2020}, and the Triplet-Singlet excitation energy (T-S) of ortho-benzyne, which is the non-diradical singlet conformer of benzyne whose excitation energy requires a proper description of dynamical correlation as well as solvent polarization~\cite{Shee2019}.
These two systems are studied in cages of water from 4 to 50 molecules (see Fig.~\ref{Fig:runtimes}) with the aim of probing collective polarization and dispersion effects. In order to avoid empirical treatment of exchange in these many-body systems, we impose a minimal distance of $\sim$ 3~\r{A} between the solute and water.

First we compute the solvation energies, $E_{\text{Solv}}\left(\text{X}\right)=E\left(\text{X}_{\text{Cage}}\right)-E\left(\text{X}_{\text{Vacuum}}\right)-E_{\text{Cage}}$, 
where $\text{X}$ is the T-shaped benzene dimer or its monomers (Table~\ref{tab:bz2_50w_cage}), in a cage composed of 50 water molecules (structures c) and d) in Fig.~\ref{Fig:runtimes}), or the ortho-benzyne in its singlet (S) and triplet (T) spin states (Table~\ref{tab:solvST}) in a cage of 4 and 30 water molecules (structures a) and b) in Fig.~\ref{Fig:runtimes}).
For all systems the El-QDO DMC results are compared with state-of-the-art DFT calculations using the PBE0 functional~\cite{ada+99jcp} augmented with the Tkatchenko-Scheffler pairwise dispersion energy (PBE0+TS)~\cite{Tkatchenko2009} or the many-body dispersion method (PBE0+MBD)~\cite{Tkatchenko2012, Ambrosetti2014, note_SI}.

\begingroup
\squeezetable
\begin{table}[!tb]
\caption{Solvation energies (in kcal/mol) of the T-shaped benzene dimer and monomers in the 50W water cage~\cite{Sirianni2022} (Molecular structures c) and d) in Fig.~\ref{Fig:runtimes}).}
\label{tab:bz2_50w_cage}
\begin{ruledtabular}
\begin{tabular}{lcccc}
                  & PBE0 & PBE0+TS & PBE0+MBD &  El-QDO DMC \\ 
    \hline
    Monomer$_1$   & -1.5 &  -4.1 & -3.8  &  -4.0(1) \\
    Monomer$_2$   & -3.4 &  -6.6 & -6.1  &  -6.4(1) \\
    Dimer         & -4.9 & -10.9 & -10.0 & -10.3(2) \\
\end{tabular}
\end{ruledtabular}
\caption{Solvation energies (in kcal/mol) of ortho-benzyne~\cite{Shee2019} in its singlet (S) and triplet (T) spin states in cages of four (4W) and thirty (30W) water molecules (Molecular structures a) and b) in Fig.~\ref{Fig:runtimes}). }\label{tab:solvST}
\label{tab:obenzyne_solvation}
\begin{ruledtabular}
\begin{tabular}{lccccc}
                  & PBE0 & PBE0+TS & PBE0+MBD &  DMC  &  El-QDO DMC  \\ 
    \hline
    S (4W)          &  -1.01  & -1.50  & -1.43 & -1.5(1)  & -1.39(5)  \\
    T (4W)          &  -0.77  & -1.24  & -1.18 & -1.0(1)  & -1.12(5)  \\
    S (30W)         &  -2.82  & -6.78  & -6.11 & -        & -6.20(7)  \\
    T (30W)         &  -2.37  & -6.41  & -5.71 & -        & -5.73(7)   \\
\end{tabular}
\end{ruledtabular}
\caption{Singlet-Triplet adiabatic excitation energies (S-T) of the ortho-benzyne~\cite{Shee2019} in vacuum (V), 4-water cage (4W) and 30-water cage (30W) (in kcal/mol), corresponding to the molecular structures a) and b) in Fig.~\ref{Fig:runtimes}. Experimental value from Ultraviolet Photoelectron Spectroscopy is estimated to be 37.5(3)~\cite{wen+98jacs}.\footnote{DMC and El-QDO DMC values are corrected with respect to the pseudopotential error shown in the SM~\cite{note_SI}}}
\label{tab:obzST}
\begin{ruledtabular}
\begin{tabular}{lccccc}
                          &  PBE0   & PBE0+TS   & PBE0+MBD   & DMC             & El-QDO DMC \\ 
      \hline
    S-T (V)                  &  28.23  & 28.23     & 28.26       & 37.23(5)        & -  \\
    S-T (4W)                 &  28.46  & 28.48     & 28.51       & 37.7(1)         & 37.50(5)  \\
    S-T (30W)                &  28.68  & 28.61     & 28.66       & -               & 37.70(8)  \\
\end{tabular}
\end{ruledtabular}
\end{table}
\endgroup

From the results reported in Table~\ref{tab:bz2_50w_cage} and Table~\ref{tab:solvST} we conclude that El-QDO DMC solvation energies are in excellent agreement with fully electronic DMC calculations (for the 4W cage for which they are feasible) and with PBE0+MBD results. 
It is known that PBE0+MBD binding energies are often quantitatively accurate for molecular systems, especially when compared to electronic DMC calculations~\cite{AlHamdani2021}. 
In fact, we note that many-body effects are particularly evident for ortho-benzyne, where the PBE0+TS method overestimates the energies when compared to the El-QDO DMC and PBE0+MBD methods, which both include, in a different manner, the many-body correlation effects of the environment on the solute. 

The key role of proper quantum embedding achieved with the El-QDO DMC approach can be observed in the singlet-triplet (S-T) adiabatic excitation energy of ortho-benzyne (Table~\ref{tab:obzST}).
Due to delocalization error, DFT severely underestimates the S-T gap~\cite{Shee2019}, while electronic DMC yields accurate results~\cite{Zhou2019} in vacuum in excellent agreement with the experimental value of 37.5(3)~kcal/mol~\cite{wen+98jacs}.
For the smallest cage of four water molecules, the excitation energies computed with electronic DMC and the El-QDO DMC approach are in excellent agreement between each other. 
Yet, the error bar of El-QDO calculations is much lower compared to electronic DMC computations for an equivalent sampling trajectory.
Moreover, as Table~\ref{tab:obzST} shows, when increasing the size of the water cage, the excitation energy changes by 0.2(1)~kcal/mol. This effect, captured by the El-QDO DMC embedding method arises from the higher polarization of the T state with respect to the S state of benzyne. We expect that future extensions of our method will allow more realistic embedding geometries where water molecules can come closer to the solute, yielding more substantial solvation effects. 
Remarkably, the computational cost of the El-QDO method is essentially independent on the number of QDOs in the embedding region (see Fig.~\ref{Fig:runtimes}) with respect to the cost of the molecule in vacuum. For the 4-water cage, the cost increases by only $\sim$1\%, and by $\sim$16\% for the 30-water cage. 
This stems from the lack of nodal structure in the QDO wavefunctions and low noise in their DMC sampling with respect to the electronic wavefunctions. 

Our study illustrates the need for fully quantum embedding methods when treating binding and excitations energies of solvated systems, and introduces a computationally efficient approach. 
The El-QDO embedding method is a firm first step in this direction as we demonstrated on the example of prototypical dispersion and hydrogen-bonded dimers and for the solvation of benzene derivates. 
A future challenge includes the incorporation of a more general representation of the QDO environment and exchange effects, which could be achieved through a synergy with machine-learning methods~\cite{mlffrev,dftbml,spookynet,paulinet}. 

\vspace{1cm}
\begin{acknowledgements}
MD and MB thank Jorge Alfonso Charry Martinez for participating in the development of the QMeCha code.
MB acknowledges financial support from the Luxembourg National Research Fund (INTER/DFG/18/12944860).
The DFT calculations presented in this paper were carried out using the HPC facilities of the University of Luxembourg~\cite{VBCG_HPCS14} {\small (see \href{http://hpc.uni.lu}{hpc.uni.lu})}.
An award of computer time was provided by the Innovative and Novel Computational Impact on Theory and Experiment (INCITE) program. This research used resources of the Argonne Leadership Computing Facility, which is a DOE Office of Science User Facility supported under Contract DE-AC02-06CH11357.
\end{acknowledgements}

\bibliography{bibliography.bib}

\end{document}